\documentclass[reprint,aps,superscriptaddress,showkeys,prl]{revtex4-2}

\usepackage{amsmath,amssymb}
\usepackage{braket}
\usepackage[hidelinks]{hyperref}
\usepackage{siunitx}
\usepackage[version=4]{mhchem}
\usepackage{todonotes}

\newcommand{\expect}[1]
{
\langle #1 \rangle
}

\renewcommand{\vec}[1]{\ensuremath{\boldsymbol{#1}}}
\newcommand{\op}[1]{\ensuremath{\hat{#1}}}
\newcommand{\avg}[1]{\ensuremath{\left\langle #1 \right\rangle}}
\newcommand{\abs}[1]{\ensuremath{\left| #1 \right|}}
\DeclareMathOperator{\tr}{tr}

\begin{document}

\title{Entanglement in Elastic Electron Scattering: Perturbation theory misses fundamental aspects of Bragg scattering.}
\author{Stefan Löffler}
\email{stefan.loeffler@tuwien.ac.at}
\affiliation{University Service Centre for Transmission Electron Microscopy, TU Wien, Stadionallee 2, 1020 Wien, Austria}
\author{Peter Schattschneider}
\affiliation{Institute of Solid State Physics, TU Wien, Wiedner Hauptstr. 8-10, 1040 Wien, Austria}

\begin{abstract}
Elastic electron scattering is one of the primary means of investigating materials on the atomic scale. It is usually described in a one-particle approach (the probe electron's evolution in a perturbative static potential), whereas we are dealing here with a two-body \emph{interaction} between the probe and the sample, both described by  separate quantum states, inducing entanglement.
In this work, we present a quantum treatment of elastic electron scattering. We find that the entanglement between probe and scatterer can have far-reaching consequences, particularly on coherence and image contrast. As a timely example, we discuss decoherence in Bragg scattering on nanoparticles. We find that conventional scattering theory is recovered in most cases.
The situation changes dramatically for freely evolving nano particles as e.g.  levitated motional ground states, an active field of research.
\end{abstract}


\maketitle

%%%%%%%%%%%%%%%%%%%%%%%%%%%%%%%%%%%%%%%%%%%%%%%%%%%%%

\section{Introduction}

Elastic scattering of charged particles is one of the primary means for investigating structural properties on the atomic scale. It's impact ranges from laying the foundation for the Rutherford model for the atom~\cite{PRSLA_v82_i557_p495,PMS6_v21_i125_p669} to imaging crystal lattices~\cite{PRSA_v236_p119} to the study of individual defects in 2D materials~\cite{NL_v8_i11_p3582} to the three-dimensional reconstruction of organic molecules such as proteins~\cite{Frank2006}.

In the present context, elastic scattering is defined as an interaction in which two interacting objects exchange only kinetic energy. The internal degrees of freedom are not involved. It is usually described as the evolution of the incident particle in the static potential of the scatterer~\cite{WilliamsCarter1996}. But strictly speaking, there is no scatterer in this standard approach which is a {\em one-particle} description in a perturbative potential whereas, more precisely, we have a Coulomb interaction between two quantum systems. The difference becomes important  when considering momentum conservation.

Take, for example, the important case of elastic scattering of an electron plane wave on a periodic lattice, leading to Bragg diffraction. Naturally, each Bragg ``beam'' corresponds to a specific momentum transfer to the probe electron. If multiple of these Bragg ``beams'' are combined coherently (as is common in electron microscopy to form a high-resolution image of the sample), the probe electron ends up in a quantum superposition of multiple momentum eigenstates. Due to momentum conservation, so must the scatterer, giving rise to interference effects. However, this violates the assumption that the scatterer is well-described by a static, fixed, perturbative potential (which has no momentum). In fact, since the groundbreaking diffraction experiment on \ce{C60} molecules~\cite{Arndt1999}, matter wave interference has been demonstrated experimentally for ever larger molecules and clusters \cite{Gerlich2011,NP_v15_p1242,ArndtArxiv} and was recently predicted for nano particles of $\sim 10^9$~amu~\cite{NimmrichterArxiv}.

In order to treat this situation, a single-particle description is evidently insufficient. Instead of a perturbative potential, the scatterer must be described as a quantum system that interacts with the incident particle. Entanglement comes into play, and consequently loss of coherence~\cite{U_v190_i_p39,JoESaRP_v241_p146810}, which is described by the reduced density matrix of the electron~\cite{Blum1996,PRB_v59_i16_p10959,M_v31_i4_p333}.

In~\cite{U_v190_i_p39}, we studied the evolution of the scatterer {\em after} interaction with the probe, and the consequences thereof for decoherence. Here, we focus on the free evolution of the scatterer {\em before} the elastic interaction, a situation, which would be expected in scattering experiments with levitated nano particles. This could be particularly relevant in the context of recent proposals predicting that scattering experiments in the TEM could be used to create superposition states of nanometer-sized particles~\cite{NimmrichterArxiv} or to entangle consecutive beam electrons~\cite{nQI_v10_p121}.

It turns out that the electron's density matrix shows signatures of increased entropy/reduced purity, in stark contrast to the one-particle description. When the mass of the scatterer tends to infinity, the differences disappear. Even for nanometer-sized scatterers, entanglement effects are  small  when free evolution is not taken into account.

%%%%%%%%%%%%%%%%%%%%%%%%%%%%%%%%%%%%%%%%%%%%%%%%%%%%%

\section{Theoretical Framework}

Throughout this work, we will use lowercase letters for denoting states in the probe beam's Hilbert space and uppercase letters for denoting states in the sample's Hilbert space.

As a measure of entanglement, we will use the density operator's purity~\cite{PR_v831_p1}, i.e. $\tr (\op{\rho}^2)$. It can easily be shown that $\tr (\op{\rho}^2) \le (\tr \op{\rho})^2 = 1$ always holds, with equality if and only if $\op{\rho}$ describes a pure state. Note that the purity is much easier to calculate than, e.g., the density operator's entropy, especially in high-dimensional Hilbert spaces.

In the following, we will assume that (i) the sample is described well by a rigid lattice with no changes to any internal degrees of freedom, thus no excitations occur and the scattering is purely elastic, (ii) the probe beam is initially in a (pure) plane-wave state while the sample is initially in a pure state, (iii) only a single elastic scattering event occurs. These assumptions are used for simplicity and clarity and are discussed further below. A general derivation can be found in the appendix.

To describe entanglement effects, it is useful to adopt the density operator formalism. Initially, the probe and sample are in independent pure states, described by the density operator $\op{\rho}_\text{in} = \ket{\vec{k}_0}\ket{I}\bra{I}\bra{\vec{k}_0}$. After the interaction (given by the Coulomb operator $\op{V}$), the system is described by $\op{\rho}_\text{tot} = \frac{1}{N_\text{tot}} \op{V} \op{\rho}_\text{in} \op{V}^\dag$ where $N_\text{tot}$ is a normalization constant. This allows for all possible outcomes of the scattering event, including inelastic scattering. Since we are only interested in elastic electron scattering, we need to project this onto all elastically scattered states
\begin{eqnarray}
\op{\rho}_\text{tot,el} = \frac{1}{N} \sum_{\vec{k},\vec{k}'} \ket{\vec{k}} \ket{\Phi_{\vec{k}}}\bra{\Phi_{\vec{k}}} \braket{\vec{k} | \op{V} | \vec{k}_0}\ket{I} \cr
\bra{I}\braket{\vec{k}_0 | \op{V}^\dag | \vec{k}'} \ket{\Phi_{\vec{k}'}}\bra{\Phi_{\vec{k}'}} \bra{\vec{k}'}
\end{eqnarray}
where the possible final states $\ket{\Phi_{\vec{k}}} = e^{-i (\vec{k}-\vec{k}_0 )\cdot \op{\bar{\vec{X}}}} \ket{I}$ are  momentum boosts of the whole lattice, i.e. an action  on the center of mass (CM) position operator $\op{\bar{\vec{X}}}$ without any changes to the internal degrees of freedom of  the initial  state $\ket{I}$ and the scattering of the electron beam from $\ket{\vec{k}_0}$ to $\ket{\vec{k}}$. For details see Eq.~\ref{eq:cPhi}.
Since the sample state is not observed (the whole point of the scattering experiment is to gain information about the sample by measuring the electron beam because the sample cannot be measured directly), one has to trace out all possible sample states, resulting in the reduced density operator of the electron in the form
\begin{equation}
\op{\rho} = \frac{1}{N} \sum_{\vec{k},\vec{k}'} c_{\vec{k}} c_{\vec{k}'}^* \braket{\Phi_{\vec{k}'} | \Phi_{\vec{k}}} \ket{\vec{k}} \bra{\vec{k}'}
\label{eq:rho}
\end{equation}
with
\begin{equation}
c_{\vec{k}} = \bra{\Phi_{\vec{k}}} \braket{\vec{k} | \op{V} | \vec{k}_0}\ket{I} = e^2 Z\sum_{j=1}^n \Braket{\Phi_{\vec{k}} | 
\frac{e^{-i (\vec{k}-\vec{k}_0) \cdot \vec{\op{X}}_j}}{\abs{\vec{k}-\vec{k}_0}^2} 
| I}
\label{eq:ck}
\end{equation}
being the matrix elements for scattering from $\ket{\vec{k}_0}\ket{I}$ to $\ket{\vec{k}}\ket{\Phi_{\vec{k}}}$ where $j$ runs over all identical $n$ sample atoms. $e$ is the electron charge, and $Z$ the atomic number.

Note that eq.~\ref{eq:rho} looks like a regular, pure state density matrix which one would expect for scattering on a fixed potential without quantum-mechanical interactions, \emph{except} for the $\braket{\Phi_{\vec{k}'} | \Phi_{\vec{k}}}$ term. If $\braket{\Phi_{\vec{k}'} | \Phi_{\vec{k}}} = 1 \ \forall \vec{k},\vec{k}'$, the density matrix is trivially separable, thus describing a pure state. Otherwise, the purity reads
\begin{equation}
    \tr\op{\rho}^2 = \frac{1}{N^2}\sum_{\vec{k},\vec{k}'} |c_{\vec{k}}|^2 \cdot |c_{\vec{k}'}|^2 \cdot |\braket{\Phi_{\vec{k}'} | \Phi_{\vec{k}}}|^2 < 1
\end{equation}
describing a mixed state as a result of entanglement. This unequivocally proves that entanglement does occur in elastic electron scattering.

It is noteworthy that in the limit of an infinitely large stationary sample, $\ket{\Phi_{\vec{k}}}$ become independent of $\vec{k}$ (since the sample does not move). In this case, $\braket{\Phi_{\vec{k}'} | \Phi_{\vec{k}}} \equiv 1$, $\op{\rho}$ becomes separable, and the results of conventional elastic scattering theory are recovered. This is not the case for finitely heavy scatterers (e.g., nanoparticles), though.

%%%%%%%%%%%%%%%%%%%%%%%%%%%%%%%%%%%%%%%%%%%%%%%%%%%%%

\section{Examples}
\subsection{Bragg scattering on Nanoparticles}

The possibility of preparing nanoparticles in a motional ground state~\cite{S_v367_p892,NP_v19_p1009,NP_v19_p1003} opens new ways for matter wave interference experiments in the electron microscope~\cite{NimmrichterArxiv}. To estimate the effect that entanglement in Bragg scattering has in such scenarios, we will consider a monoatomic lattice. Under the assumption of a rigid lattice described above, $\ket{\Phi_{\vec{k}}} = e^{-i \vec{q} \cdot \op{\bar{\vec{X}}}} \ket{I}$, where $\vec{q} = \vec{k} - \vec{k}_0$ is the momentum transfer and $\op{\bar{\vec{X}}} = \frac{1}{n} \sum_{j=1}^n \op{\vec{X}}_j$ is the position operator of the center of mass (CM) of a rigid lattice with $n$ identical atoms. With these assumptions, the quantum central limit theorem \cite{E_v25_p600} is applicable (see appendix)
\begin{equation}
\op{\rho} \to \frac{1}{N} \sum_{\vec{q}, \vec{q}'} e^{-(q^2 + {q'}^2 - \vec{q}'\cdot\vec{q}) \sigma_0^2} \cdot
f_{\vec{q}} f_{\vec{q}'}^* \ket{\vec{k}_0 + \vec{q}}\bra{\vec{k}_0 + \vec{q}'}
\label{eq:rho-asymptotic}
\end{equation}
where $\sigma_0$ is the standard deviation of the CM probability distribution and the $f_{\vec{q}}$ are the conventional Bragg scattering amplitudes. Here, $q$ is typically of the order of \qty{10}{\per\nano\meter} (for small-angle scattering).

Delic et al.~\cite{S_v367_p892} report laser cooling of a silica nanoparticle of $\sim \qty{10}{\nano\meter}$ comprising $\sim 10^8$ atoms to the motional ground state with a CM localization of $\sigma_0 \sim \qty{3}{\pico\meter}$. This  puts the exponential factor in the regime of $0.99$, so $\op \rho$ is almost indistinguishable from the density matrix of a pure state.
This changes dramatically when dispersion of the sample wavefunction is taken into account, which occurs naturally between interactions. Assuming Gaussian-like dispersion, the variance at time $t$ is given by~\cite{CohenTannoudji2020}
\begin{equation}
    \sigma_0^2 \mapsto \sigma_0^2 \left(1 + \left( \frac{\hbar  t}{M\sigma_0^2}\right)^2 \right)
\end{equation}
where $M$ is the total mass of the scatterer. This results in a decoherence time $\tau$ (after which the exponential factor for the probe beam has decayed to $1/e$)  of
\begin{equation}
    \tau = \frac{\sigma_0^2 M}{\hbar} \sqrt{\frac{1}{q^2 \sigma_0^2} - 1}.
\end{equation}
It is evident that for $M \rightarrow \infty$, there is no decoherence, and the standard approach to Bragg scattering is recovered. However, if the same localization precision as in \cite{S_v367_p892} is feasible for $M=\qty{720}{amu}$ (a $C_{60}$ fullerene), $\tau$ resolves to $\approx\qty{3}{\pico\second}$. For heavier samples (all else being equal), it resolves to $\approx\qty{5}{\nano\second}$ ($M=\qty{1e6}{amu}$) and $\approx\qty{9}{\micro\second}$ ($M=\qty{2e9}{amu}$~\cite{S_v367_p892}).
Of course, this can only be seen in the case of free sample dispersion and if no other decoherence mechanisms (such as interaction with the environment) affect the probe beam on a shorter timescale.

%%%%%%%%%%%%%%%%%%%%%%%%%%%%%%%%%%%%%%%%%%%%%%%%%%%%%

\subsection{Testing the Purity}
Consider the fringe contrast obtained from the post selection of two Bragg reflections  $\vec{q} \in \{ -\vec{G}, \vec{G} \}$ and $f_{-\vec{G}} = f_{\vec{G}} = f$. This can be realized with contrast apertures blocking the direct beam and all other Bragg reflections~\cite{U_v190_i_p39}. Then,  eq.~\ref{eq:rho-asymptotic} reads in a $\{\ket{\vec{k}_0 - \vec{G}}, \ket{\vec{k}_0 + \vec{G}}\}$ basis
\begin{equation}
\rho_k = \frac{1}{2} \begin{pmatrix}
1 & e^{-2G^2 \sigma_0^2} \\
e^{-2G^2 \sigma_0^2} & 1
\end{pmatrix}
\end{equation}
Note that the scattering amplitudes $f_{\mp \vec{G}}$ do not appear in the  density matrix  because they can be extracted as a prefactor that disappears when  normalizing $\rho_k$ to 1.
For this 2-dimensional density matrix, it is straight-forward to calculate the purity 
\begin{equation}
    \tr \rho_k^2 = \frac{1}{2}(1 + e^{-4G^2 \sigma_0^2})
\end{equation} as well as the von Neumann entropy 
\begin{multline}
    -\tr[\rho_k\log\rho_k] = \log 2 - \frac{1}{2} \left[ (1 + e^{-4G^2 \sigma_0^2}) \log(1 + e^{-4G^2 \sigma_0^2}) \right. \\
    \left. + (1 - e^{-4G^2 \sigma_0^2})\log(1 - e^{-4G^2 \sigma_0^2})\right]
\end{multline}
The fringe contrast is given by the real-space diagonal elements
\begin{equation}
I(\vec{r}) = \braket{\vec{r} | \op{\rho} | \vec{r}} = 
1 + e^{-2G^2 \sigma_0^2} \cos(2\vec{G} \cdot \vec{r})
\end{equation}
Thus, the contrast of the lattice fringes in the image depends directly on the ``entanglement term'' $e^{-G^2 \sigma_0^2}$ as
\begin{equation}
\frac{I_\text{max} - I_\text{min}}{I_\text{max} + I_\text{min}} = e^{-2G^2 \sigma_0^2}
\end{equation}

%%%%%%%%%%%%%%%%%%%%%%%%%%%%%%%%%%%%%%%%%%%%%%%%%%%%%

\section{Discussion}
Multiple assumptions and approximations went into the derivation of the theoretical framework that warrant closer inspection.  It should be realized, however, that removing the three most important approximations discussed in the following tend to decrease the purity of the probe further. What we have shown here is that even in an idealized case, entanglement reduces the purity and introduces entropy. 

The first assumption was that the sample is described well by a rigid lattice and no excitations occur. Strictly speaking, this is not true as many inelastic channels exist (e.g., plasmons and core-level excitations). With suitable monochromators and energy filters, those can be excluded from experiments. State-of-the art electron microscopes reach energy resolutions $<\qty{10}{\milli\electronvolt}$~\cite{N_v514_p209}, eliminating the dominant contributions to inelastic scattering. Some ultra-low energy excitations may remain, but they are usually many orders of magnitude weaker than the elastic peak. Additionally, they can be reduced further by cooling the sample \cite{S_v367_p892,NP_v19_p1009,NP_v19_p1003}.
Without worrying about internal excitations, the description of the sample essentially boils down to its center-of-mass, with all atoms receiving the same elastic momentum ``boost''.

The second assumption was that both the probe beam and the sample are initially in a pure state. This is perhaps the most severe simplification, especially for the sample. Such a state could potentially be realized by capturing the sample in a optical trap and cooling it to the ground state. However, it is equally possible to express the density operator describing a mixed state as an incoherent sum of multiple density operators each corresponding to a pure state and weighted by a suitable probability factor. Then, the theoretical framework developed here can be applied to each pure-state density operator separately. The general result is the same, however, and was therefore omitted here for brevity and clarity.

The third assumption was that only a single elastic scattering occurs. Again, this is certainly not the case for samples thicker than a monolayer. However, algorithms exist to generalize single scattering to many atoms, first and foremost the multislice algorithm~\cite{Kirkland1998,U_v131_i0_p39,PRB_v96_p245121a}. As this treatment is well-known and does not add anything to the fundamental discussion of entanglement in elastic scattering, it was omitted here as well.

Without taking dispersion of the scatterer into account, the effect of entanglement is present, but typically small. This changes dramatically, however, when considering scatterers in delocalized quantum states, e.g., after dispersing freely for even a few picoseconds. In that case, the effect of entanglement becomes very noticable or even dominant. Free dispersions occurs when the sample is not interacting with the environment, i.e., when decoherence is avoided. Many possibilities exist to achieve that, from dropping the sample initially levitating in a optical trap~\cite{Bateman2014,PRL_v107_p20405,Rossi2024} to engineering the environment (e.g., better vacuum, reduced thermal radiation, etc.)~\cite{PR_v831_p1}.

%%%%%%%%%%%%%%%%%%%%%%%%%%%%%%%%%%%%%%%%%%%%%%%%%%%%%

\section{Conclusion \& Outlook}
From the above, it becomes clear that even when including entanglement into the model, for the majority of electron microscopy measurements and simulations we  may call off the alarm: We may safely eliminate doubts about what legions of microscopists have been doing over a century now. Although under normal conditions, entanglement and loss of coherence induce only miniscule effects, they are observable, even dominant when the scatterer is allowed to evolve freely before interacting with the probing electron. To this aim, one could consider the preparation of nano particles in a levitated motional ground state, as recently achieved~\cite{S_v367_p892,N_v595_p378,NP_v21_p1603,ArndtArxiv2}. Such experiments will facilitate novel diffraction schemes and shed new light on the fundamentals of elastic scattering.

\begin{acknowledgments}
The authors thank Ph. Haslinger, S. Nimmrichter and D. Rätzel for fruitful discussions. The authors acknowledge TU Wien Bibliothek for financial support through its Open Access Funding Programme.
\end{acknowledgments}

%%%%%%%%%%%%%%%%%%%%%%%%%%%%%%%%%%%%%%%%%%%%%%%%%%%%%


\appendix

\section{General Derivation of the Reduced Density Operator After Scattering}\label{sec:general-derivation}

Suppose the total system (comprising the probe and the sample sub-systems) is initially described by the density operator $\op{\rho}_\text{in}$.
Then the total density operator after the interaction (described by the interaction operator $\op{V}$) is
\begin{equation}
    \op{\rho}_\text{tot} = \frac{1}{N} \op{V} \op{\rho}_\text{in} \op{V}^\dag
    \quad \text{with} \quad N = \tr \op{V} \op{\rho}_\text{in} \op{V}^\dag
\end{equation}
The probe beam's reduced density operator thus reads
\begin{equation}
    \op{\rho} = \frac{1}{N} \sum_F \braket{F | \op{V} \op{\rho}_\text{in} \op{V}^\dag | F}
\end{equation}
with the purity
\begin{equation}
    \tr_{f} \op{\rho}^2
    = \frac{1}{N^2} \sum_{f,f'} \left| \sum_{F} \bra{f} \braket{F | \op{V} \op{\rho}_\text{in} \op{V}^\dag | F} \ket{f'}\right|^2
\end{equation}
Expanded in a $\{\ket{f}\}$ orthonormal basis, the reduced density operator reads $\op{\rho} = \sum_{f,f'} \gamma_{f,f'} \ket{f}\bra{f'}$
with
\begin{equation}
\label{eq:gammaf}
\gamma_{f,f'} = \frac{1}{N} \sum_F \bra{f} \braket{F | \op{V} \op{\rho}_\text{in} \op{V}^\dag | F} \ket{f'}.
\end{equation}
Clearly
\begin{equation}
\tr\op{\rho}^2 = \sum_{f,f'} |\gamma_{f,f'}|^2
\end{equation}

The derivation above is fully general and allows for any kind of scattering, including inelastic scattering. In this work, we are only interested in elastic scattering. This can be incorporated by replacing $\ket{F} \mapsto \sum_{\Phi_f} \ket{\Phi_f}\braket{\Phi_f | F}$ where the sum runs over all sample states $\{ \ket{\Phi_f} \}$ compatible with elastic scattering to $\ket{f}$ (and similarly for scattering to $\ket{f'}$).
This leads to 
\begin{gather}
\gamma_{f,f'} = \frac{1}{N} \sum_F \sum_{\Phi_{f},\Phi'_{f'}} \braket{F | \Phi_{f}} \bra{\Phi_{f}} \braket{f | \op{V} \op{\rho}_\text{in} \op{V}^\dag | f'} \ket{\Phi'_{f'}} \braket{\Phi'_{f'} | F} \notag\\
= \frac{1}{N} \sum_{\Phi_{f},\Phi'_{f'}} \braket{\Phi'_{f'} | \Phi_{f}} \bra{\Phi_{f}} \braket{f | \op{V} \op{\rho}_\text{in} \op{V}^\dag | f'} \ket{\Phi'_{f'}}
\end{gather}

Assuming an initial pure state $\op{\rho}_\text{in} = \ket{I}\ket{i}\bra{i}\bra{I}$ yields
$\gamma_{f,f'} = \frac{1}{N} \sum_{\Phi_{f},\Phi'_{f'}} c_{\Phi_{f}} c_{\Phi'_{f'}}^*\braket{\Phi'_{f'} | \Phi_{f}}$
with
\begin{equation}
	c_{\Phi_{f}} = \bra{\Phi_{f}} \braket{f | \op{V} | i} \ket{I}
    \label{eq:cPhi}
\end{equation}

Further assuming plane wave Coulomb scattering (i.e., $\ket{i} = \ket{\vec{k}_0}$, $\ket{f} = \ket{\vec{k}_0 + \vec{q}}$) and $\ket{\Phi_f} =  e^{-i \vec{q} \cdot \op{\bar{\vec{X}}}} \ket{I}$ (i.e., only considering a single final sample state that is a momentum boost of the whole lattice without any changes to its internal state) gives
\begin{gather}
\gamma_{f,f'} = \frac{1}{N} c_{\vec{q}} c_{\vec{q}'}^*\braket{I | e^{i (\vec{q}' - \vec{q}) \cdot \op{\bar{\vec{X}}}} | I} \\
c_{\vec{q}} = \frac{Z e^2}{q^2} \braket{I | \sum_{j=1}^n e^{i \vec{q} \cdot (\op{\bar{\vec{X}}} - \op{\vec{X}}_j)} | I}
\end{gather}

\section{Large Mono-atomic Samples}
\label{sec:general-qclt}
Here, we will consider the behavior of asymptotically large (number of atoms $n \to \infty$) samples that consist of non-interacting, identical atoms at positions $\vec{R}_j$. In this model, $\ket{I} = \otimes_{j=1}^n \ket{\phi_j}$ and $\braket{\vec{X}_j | \phi_j} = \phi(\vec{X}_j-\vec{R}_j)$, i.e. we have essentially one atomic wavefunction $\phi$, from which the wavefunction $\phi_j$ of the $j$-th atom is generated by shifting, and the total initial state $\ket{I}$ is given by the direct product of all these shifted atomic wavefunctions.

Then, the quantum central limit theorem \cite{E_v25_p600} is applicable which states that for a general function $f$, the following holds asymptotically:
\begin{multline}
\lim_{n \to\infty} \Braket{I | f\left( \frac{1}{n} \sum_{i}^n \op{\vec{x}}_i\right) | I } \\
= \frac{1}{(\frac{\sigma_x}{\sqrt{n}} (2\pi)^{3/2})^3} \int d\vec{\chi} f(\vec{\chi}) e^{-\frac{(\vec{\chi} - \avg{\vec{x}})^2}{2(\frac{\sigma_x}{\sqrt{n}})^2}}
\end{multline}
with $\avg{\vec{x}} = \braket{\phi|\vec{x}|\phi}$, i.e., the expectation value of the position operator w.r.t. a single independent atom, $\sigma_x^2 = \avg{x^2} - \avg{x}^2$, i.e., the variance of the position w.r.t. a single independent atom.

With $\op{\bar{\vec{X}}} = \frac{1}{n} \sum_{i}^n \op{\vec{x}}_i$, $f(\vec{\chi}) = e^{i (\vec{q}'-\vec{q}) \cdot \vec{\chi}}$, and $\sigma_0 = \sigma_x / \sqrt{n}$, we get
\[
    \braket{I | e^{i (\vec{q}'-\vec{q}) \cdot \op{\bar{\vec{X}}}} | I} \to
    \frac{1}{(\sigma_0 (2\pi)^{3/2})^3} \int d\vec{\chi} e^{i (\vec{q}'-\vec{q}) \cdot \vec{\chi}} e^{-\frac{(\vec{\chi} - \avg{\vec{x}})^2}{2\sigma_0^2}}
\]
which is the Fourier-transform of a Gaussian shifted by $\avg{\vec{x}}$ and results in
\[
    \braket{I | e^{i (\vec{q}'-\vec{q}) \cdot \op{\bar{\vec{X}}}} | I} \to
    \frac{1}{(2\pi)^3} e^{i (\vec{q}'-\vec{q}) \cdot \avg{\vec{x}}} e^{-\frac{(\vec{q}'-\vec{q})^2 \sigma_0^2}{2}}.
\]
With $\op{\vec{x}} = \op{\vec{X}}_j - \vec{R}_j$, we additionally have
\begin{gather*}
\avg{\op{\vec{x}}} = \frac{1}{n} \sum_i^n \avg{\op{\vec{x}}} = \frac{1}{n} \sum_i^n (\avg{\op{\vec{X}}_i} - \vec{R}_j)
\\
= \avg{\op{\bar{\vec{X}}}} - \frac{1}{n} \sum_i^n \vec{R}_i
= \braket{I | \op{\bar{\vec{X}}} | I} - \vec{\bar{R}}
\end{gather*}
where $\vec{\bar{R}} = \frac{1}{n} \sum_i^n \vec{R}_i$ is the center-of-mass of the entire, large sample. Thus, $\gamma_{f,f'}$ asymptotically approaches the form
\[
\gamma_{f,f'} \to \frac{1}{N (2\pi)^3} c_{\vec{q}} c_{\vec{q}'}^* e^{i(\vec{q}'-\vec{q}) \cdot (\braket{I | \op{\bar{\vec{X}}}| I} - \vec{\bar{R}})} e^{-(\vec{q}'-\vec{q})^2 \frac{\sigma_0^2}{2}}
\]

For $c_{\vec{q}}$, a similar computation can be performed. For a specific index $j$ and recalling the definition $\ket{I} = \otimes_i^n \ket{\phi_j}$, we get
\begin{gather*}
    \braket{I | e^{i \vec{q} \cdot ((\frac{1}{n} \sum_i^n \op{\vec{X}}_i) - \op{\vec{X}}_j)} | I} \\
    = \braket{\phi_j | e^{i \vec{q} \cdot (\frac{1}{n} - 1) \op{\vec{X}}_j} | \phi_j} \prod_{i\ne j}^n \braket{\phi_i | e^{i \vec{q} \cdot (\frac{1}{n} \op{\vec{X}}_i)} | \phi_i} \\
    = \frac{\braket{\phi_j | e^{i \vec{q} \cdot (\frac{1}{n} - 1) \op{\vec{X}}_j} | \phi_j}}{\braket{\phi_j | e^{i \vec{q} \cdot (\frac{1}{n} \op{\vec{X}}_j)} | \phi_j}} \prod_i^n \braket{\phi_i | e^{i \vec{q} \cdot (\frac{1}{n} \op{\vec{X}}_i)} | \phi_i}
\end{gather*}
The second term is exactly $\braket{I | e^{i \vec{q} \cdot \op{\bar{\vec{X}}}} | I}$, to which the quantum central limit theorem can be applied. To simplify the first term, we exploit the fact that all atomic wave-functions are assumed to be shifted copies of each other (i.e., $\op{\vec{x}} = \op{\vec{X}}_j - \vec{R}_j$), giving
\begin{gather*}
    \frac{\braket{\phi_j | e^{i \vec{q} \cdot (\frac{1}{n} - 1) \op{\vec{X}}_j} | \phi_j}}{\braket{\phi_j | e^{i \vec{q} \cdot (\frac{1}{n} \op{\vec{X}}_j)} | \phi_j}}
    =
    \frac{\braket{\phi | e^{i \vec{q} \cdot (\frac{1}{n} - 1) (\op{\vec{x}} + \vec{R}_j)} | \phi}}{\braket{\phi | e^{i \vec{q} \cdot \frac{1}{n} (\op{\vec{x}} + \vec{R}_j)} | \phi}}
    \\
    = \frac{\braket{\phi | e^{i \vec{q} \cdot (\frac{1}{n} - 1) \op{\vec{x}}} | \phi} e^{i\vec{q} \cdot (\frac{1}{n}-1)\vec{R}_j}}{\braket{\phi | e^{i \vec{q} \cdot \frac{1}{n} \op{\vec{x}}} | \phi} e^{i\vec{q} \cdot \frac{1}{n}\vec{R}_j}}
    \\
    = \frac{\braket{\phi | e^{i \vec{q} \cdot (\frac{1}{n} - 1) \op{\vec{x}}} | \phi}}{\braket{\phi | e^{i \vec{q} \cdot \frac{1}{n} \op{\vec{x}}} | \phi}}  e^{-i\vec{q} \cdot \vec{R}_j}
    \\
    \to \braket{\phi | e^{-i \vec{q} \cdot \op{\vec{x}}} | \phi} e^{-i\vec{q} \cdot \vec{R}_j}
\end{gather*}
where the last line is again the asymptotic behavior for large $n$.

Putting it all together, we find for $c_{\vec{q}}$ the asymptotic behavior of
\begin{gather*}
    c_{\vec{q}} \to \frac{Ze^2}{q^2 (2\pi)^3} \sum_j^n \braket{\phi | e^{-i \vec{q} \cdot \op{\vec{x}}} | \phi} e^{-i\vec{q} \cdot \vec{R}_j}
    e^{i\vec{q}\cdot\avg{\vec{x}}} e^{-q^2\frac{\sigma_0^2}{2}} \\
    =
    \frac{Ze^2}{q^2 (2\pi)^3} \braket{\phi | e^{-i \vec{q} \cdot \op{\vec{x}}} | \phi}
    e^{i\vec{q}\cdot\avg{\vec{x}}} e^{-q^2\frac{\sigma_0^2}{2}} \sum_j^n e^{-i\vec{q} \cdot \vec{R}_j} \\
\end{gather*}
As above, we use $\avg{\op{\vec{x}}} = \braket{I | \op{\bar{\vec{X}}} | I} - \vec{\bar{R}}$. This leads to the final result
\begin{gather}
c_{\vec{q}} \to e^{-q^2 \frac{\sigma_0^2}{2}} e^{i \vec{q} \cdot (\braket{I | \op{\bar{\vec{X}}}| I} - \vec{\bar{R}})} \cdot f_{\vec{q}}\\
f_{\vec{q}} = \frac{Z e^2}{q^2 (2\pi)^3} 
\braket{\phi | e^{-i \vec{q} \cdot \op{\vec{x}}} | \phi}
\sum_{j=1}^n e^{-i \vec{q} \cdot \vec{R}_j}
\end{gather}
Note that the same result is obtained for arbitrary (even small) $n$ if the initial wavefunction can be described by independent Gaussians.
Clearly, the $c_{\vec{q}}$ are the Fourier representations of the convolution of an entanglement-based damping factor (first term) with the conventional scattering amplitude $f_{\vec{q}}$ (comprising the Rutherford envelope, the charge distribution of a single atom, and the delta comb of the atom positions).

Reinserting the $c_{\vec{q}}$ into $\gamma_{f,f'}$, some phase factors cancel and we get
\begin{equation}
\gamma_{f,f'} \to \frac{1}{N (2\pi)^3} e^{-(q^2 + {q'}^2 - \vec{q}'\cdot\vec{q}) \sigma_0^2} \cdot f_{\vec{q}} f_{\vec{q}'}^*
\end{equation}
Again, the first factor is the entanglement factor, while the rest comprises the conventional scattering amplitudes.

\section{Simplified derivation of the QCLT}\label{sec:cltd}
Here, we give a simplified derivation of the quantum central limit theorem (QCLT). We assume an arrangement of $n \gg 1$ identical, independent atoms, shifted by $R_j$ from the atom centered at the origin, $\expect{x}=\int x |\phi(x)|^2 dx=0$. For easier readability, we discard the bold face notation for 3-vectors. 

The matrix element  in Eq.~\ref{eq:rho} is 
\[
\braket{\Phi_{k}|\Phi_{k'}}=\Pi_j^n \int |\phi_j(X_j)|^2 e^{iq X_j/n} \, dX_j . 
\]
where we abbreviate $q=(k-k')$ and  $\phi_j(X_j)=\phi(X_j-R_j)$. Substitution $X_j-R_j \mapsto  x$ gives
\begin{eqnarray}
\braket{\Phi_{k}|\Phi_{k'}}&=&e^{iq \sum_j^n R_j/n} \Pi_j^n \int |\phi(x)|^2 e^{iq x/n} \, dx \cr
&=&
e^{iq \sum_j^n R_j/n} [\tilde\varphi(q/n)]^n. 
\end{eqnarray}
where we defined 
\[
\tilde\varphi(q)=\int |\phi(x)|^2 e^{iq x} \, dx.
\]
The rightmost expression follows from the fact that all factors in the product are equal. 
$\tilde \varphi$ is known as the characteristic function of the probability distribution $|\phi|^2$.~\footnote{\url{https://en.wikipedia.org/wiki/Characteristic_function_(probability_theory)}, accessed March 6th, 2026}.
 Expanding $\tilde\varphi$ into a Taylor series up to second order
\[
\tilde\varphi(q/n)=\tilde\varphi(0) + \tilde\varphi'(0) \frac{q}{n} + \tilde\varphi''(0) \frac{q^2}{2 n^2}
\]
with the $k$-th derivative 
\[
\tilde \varphi^{(k)}(0)=i^k \expect{x^k}.
\]
Then 
\begin{eqnarray}
\braket{\Phi_{k}|\Phi_{k'}} &\doteq& e^{iq \sum_j^n R_j/n} [1 + i\expect{
x} \frac{q}{n} - \expect{x^2} \frac{q^2}{2 n^2}]^n \cr 
&=&e^{iq \sum_j^nR_j/n} \left[1 - \frac{\expect{x^2} q^2}{2 n}\frac{1}{n} \right]^n
\end{eqnarray}
because $\expect{x}=0$. Observing that for large $n$ the last term tends to the exponential function, it follows
\[
  \braket{\Phi_{k}|\Phi_{k'}} \doteq  e^{iq \bar R} e^{-\frac{\sigma_0 q^2}{2}}
\]
where  $\bar R=\sum_j^nR_j/n$ is  the coordinate of the center of mass, and $\sigma_0=\sqrt{\expect{x^2}/{n}}$. 

This approach works also for the coefficients Eq.~\ref{eq:ck}, as shown in the section
 on large samples above. 

\begin{thebibliography}{34}%
\makeatletter
\providecommand \@ifxundefined [1]{%
 \@ifx{#1\undefined}
}%
\providecommand \@ifnum [1]{%
 \ifnum #1\expandafter \@firstoftwo
 \else \expandafter \@secondoftwo
 \fi
}%
\providecommand \@ifx [1]{%
 \ifx #1\expandafter \@firstoftwo
 \else \expandafter \@secondoftwo
 \fi
}%
\providecommand \natexlab [1]{#1}%
\providecommand \enquote  [1]{``#1''}%
\providecommand \bibnamefont  [1]{#1}%
\providecommand \bibfnamefont [1]{#1}%
\providecommand \citenamefont [1]{#1}%
\providecommand \href@noop [0]{\@secondoftwo}%
\providecommand \href [0]{\begingroup \@sanitize@url \@href}%
\providecommand \@href[1]{\@@startlink{#1}\@@href}%
\providecommand \@@href[1]{\endgroup#1\@@endlink}%
\providecommand \@sanitize@url [0]{\catcode `\\12\catcode `\$12\catcode
  `\&12\catcode `\#12\catcode `\^12\catcode `\_12\catcode `\%12\relax}%
\providecommand \@@startlink[1]{}%
\providecommand \@@endlink[0]{}%
\providecommand \url  [0]{\begingroup\@sanitize@url \@url }%
\providecommand \@url [1]{\endgroup\@href {#1}{\urlprefix }}%
\providecommand \urlprefix  [0]{URL }%
\providecommand \Eprint [0]{\href }%
\providecommand \doibase [0]{https://doi.org/}%
\providecommand \selectlanguage [0]{\@gobble}%
\providecommand \bibinfo  [0]{\@secondoftwo}%
\providecommand \bibfield  [0]{\@secondoftwo}%
\providecommand \translation [1]{[#1]}%
\providecommand \BibitemOpen [0]{}%
\providecommand \bibitemStop [0]{}%
\providecommand \bibitemNoStop [0]{.\EOS\space}%
\providecommand \EOS [0]{\spacefactor3000\relax}%
\providecommand \BibitemShut  [1]{\csname bibitem#1\endcsname}%
\let\auto@bib@innerbib\@empty
%</preamble>
\bibitem [{\citenamefont {Geiger}\ and\ \citenamefont
  {Marsden}(1909)}]{PRSLA_v82_i557_p495}%
  \BibitemOpen
  \bibfield  {author} {\bibinfo {author} {\bibfnamefont {H.}~\bibnamefont
  {Geiger}}\ and\ \bibinfo {author} {\bibfnamefont {E.}~\bibnamefont
  {Marsden}},\ }\bibfield  {title} {\bibinfo {title} {On a diffuse reflection
  of the $\alpha$-particles},\ }\href {https://doi.org/10.1098/rspa.1909.0054}
  {\bibfield  {journal} {\bibinfo  {journal} {Proc. R. Soc. Lond. A}\ }\textbf
  {\bibinfo {volume} {82}},\ \bibinfo {pages} {495} (\bibinfo {year}
  {1909})}\BibitemShut {NoStop}%
\bibitem [{\citenamefont {Rutherford}(1911)}]{PMS6_v21_i125_p669}%
  \BibitemOpen
  \bibfield  {author} {\bibinfo {author} {\bibfnamefont {E.}~\bibnamefont
  {Rutherford}},\ }\bibfield  {title} {\bibinfo {title} {The scattering of
  $\alpha$ and $\beta$ particles by matter and the structure of the atom},\
  }\href {https://doi.org/10.1080/14786440508637080} {\bibfield  {journal}
  {\bibinfo  {journal} {Philosophical Magazine. Series 6}\ }\textbf {\bibinfo
  {volume} {21}},\ \bibinfo {pages} {669} (\bibinfo {year} {1911})}\BibitemShut
  {NoStop}%
\bibitem [{\citenamefont {Menter}(1956)}]{PRSA_v236_p119}%
  \BibitemOpen
  \bibfield  {author} {\bibinfo {author} {\bibfnamefont {J.~W.}\ \bibnamefont
  {Menter}},\ }\bibfield  {title} {\bibinfo {title} {The direct study by
  electron microscopy of crystal lattices and their imperfections},\ }\href
  {https://doi.org/10.1098/rspa.1956.0117} {\bibfield  {journal} {\bibinfo
  {journal} {Proc. Roy. Soc. A}\ }\textbf {\bibinfo {volume} {236}},\ \bibinfo
  {pages} {119} (\bibinfo {year} {1956})}\BibitemShut {NoStop}%
\bibitem [{\citenamefont {Meyer}\ \emph {et~al.}(2008)\citenamefont {Meyer},
  \citenamefont {Kisielowski}, \citenamefont {Erni}, \citenamefont {Rossell},
  \citenamefont {Crommie},\ and\ \citenamefont {Zettl}}]{NL_v8_i11_p3582}%
  \BibitemOpen
  \bibfield  {author} {\bibinfo {author} {\bibfnamefont {J.~C.}\ \bibnamefont
  {Meyer}}, \bibinfo {author} {\bibfnamefont {C.}~\bibnamefont {Kisielowski}},
  \bibinfo {author} {\bibfnamefont {R.}~\bibnamefont {Erni}}, \bibinfo {author}
  {\bibfnamefont {M.~D.}\ \bibnamefont {Rossell}}, \bibinfo {author}
  {\bibfnamefont {M.~F.}\ \bibnamefont {Crommie}},\ and\ \bibinfo {author}
  {\bibfnamefont {A.}~\bibnamefont {Zettl}},\ }\bibfield  {title} {\bibinfo
  {title} {Direct imaging of lattice atoms and topological defects in graphene
  membranes},\ }\href {https://doi.org/10.1021/nl801386m} {\bibfield  {journal}
  {\bibinfo  {journal} {Nano Lett.}\ }\textbf {\bibinfo {volume} {8}},\
  \bibinfo {pages} {3582} (\bibinfo {year} {2008})}\BibitemShut {NoStop}%
\bibitem [{\citenamefont {Frank}(2006)}]{Frank2006}%
  \BibitemOpen
  \bibfield  {author} {\bibinfo {author} {\bibfnamefont {J.}~\bibnamefont
  {Frank}},\ }\href {https://doi.org/10.1093/acprof:oso/9780195182187.001.0001}
  {\emph {\bibinfo {title} {Three-Dimensional Electron Microscopy of
  Macromolecular Assemblies}}}\ (\bibinfo  {publisher} {Oxford University
  Press},\ \bibinfo {year} {2006})\BibitemShut {NoStop}%
\bibitem [{\citenamefont {Williams}\ and\ \citenamefont
  {Carter}(1996)}]{WilliamsCarter1996}%
  \BibitemOpen
  \bibfield  {author} {\bibinfo {author} {\bibfnamefont {D.~B.}\ \bibnamefont
  {Williams}}\ and\ \bibinfo {author} {\bibfnamefont {C.~B.}\ \bibnamefont
  {Carter}},\ }\href {https://doi.org/10.1007/978-1-4757-2519-3} {\emph
  {\bibinfo {title} {Transmission electron microscopy}}}\ (\bibinfo
  {publisher} {Plenum Press},\ \bibinfo {address} {New York},\ \bibinfo {year}
  {1996})\BibitemShut {NoStop}%
\bibitem [{\citenamefont {Arndt}\ \emph {et~al.}(1999)\citenamefont {Arndt},
  \citenamefont {Nairz}, \citenamefont {Vos-Andreae}, \citenamefont {Keller},
  \citenamefont {van~der Zouw},\ and\ \citenamefont {Zeilinger}}]{Arndt1999}%
  \BibitemOpen
  \bibfield  {author} {\bibinfo {author} {\bibfnamefont {M.}~\bibnamefont
  {Arndt}}, \bibinfo {author} {\bibfnamefont {O.}~\bibnamefont {Nairz}},
  \bibinfo {author} {\bibfnamefont {J.}~\bibnamefont {Vos-Andreae}}, \bibinfo
  {author} {\bibfnamefont {C.}~\bibnamefont {Keller}}, \bibinfo {author}
  {\bibfnamefont {G.}~\bibnamefont {van~der Zouw}},\ and\ \bibinfo {author}
  {\bibfnamefont {A.}~\bibnamefont {Zeilinger}},\ }\bibfield  {title} {\bibinfo
  {title} {Wave-particle duality of \ce{C60} molecules},\ }\href
  {https://doi.org/10.1038/44348} {\bibfield  {journal} {\bibinfo  {journal}
  {Nature}\ }\textbf {\bibinfo {volume} {401}},\ \bibinfo {pages} {680}
  (\bibinfo {year} {1999})}\BibitemShut {NoStop}%
\bibitem [{\citenamefont {Gerlich}\ \emph {et~al.}(2011)\citenamefont
  {Gerlich}, \citenamefont {Eibenberger}, \citenamefont {Tomandl},
  \citenamefont {Nimmrichter}, \citenamefont {Hornberger}, \citenamefont
  {Fagan}, \citenamefont {Tüxen}, \citenamefont {Mayor},\ and\ \citenamefont
  {Arndt}}]{Gerlich2011}%
  \BibitemOpen
  \bibfield  {author} {\bibinfo {author} {\bibfnamefont {S.}~\bibnamefont
  {Gerlich}}, \bibinfo {author} {\bibfnamefont {S.}~\bibnamefont
  {Eibenberger}}, \bibinfo {author} {\bibfnamefont {M.}~\bibnamefont
  {Tomandl}}, \bibinfo {author} {\bibfnamefont {S.}~\bibnamefont
  {Nimmrichter}}, \bibinfo {author} {\bibfnamefont {K.}~\bibnamefont
  {Hornberger}}, \bibinfo {author} {\bibfnamefont {P.~J.}\ \bibnamefont
  {Fagan}}, \bibinfo {author} {\bibfnamefont {J.}~\bibnamefont {Tüxen}},
  \bibinfo {author} {\bibfnamefont {M.}~\bibnamefont {Mayor}},\ and\ \bibinfo
  {author} {\bibfnamefont {M.}~\bibnamefont {Arndt}},\ }\bibfield  {title}
  {\bibinfo {title} {Quantum interference of large organic molecules},\ }\href
  {https://doi.org/10.1038/ncomms1263} {\bibfield  {journal} {\bibinfo
  {journal} {Nature Comm}\ }\textbf {\bibinfo {volume} {2}} (\bibinfo {year}
  {2011})}\BibitemShut {NoStop}%
\bibitem [{\citenamefont {Fein}\ \emph {et~al.}(2019)\citenamefont {Fein},
  \citenamefont {Geyer}, \citenamefont {Zwick}, \citenamefont {Kiałka},
  \citenamefont {Pedalino}, \citenamefont {Mayor}, \citenamefont {Gerlich},\
  and\ \citenamefont {Arndt}}]{NP_v15_p1242}%
  \BibitemOpen
  \bibfield  {author} {\bibinfo {author} {\bibfnamefont {Y.~Y.}\ \bibnamefont
  {Fein}}, \bibinfo {author} {\bibfnamefont {P.}~\bibnamefont {Geyer}},
  \bibinfo {author} {\bibfnamefont {P.}~\bibnamefont {Zwick}}, \bibinfo
  {author} {\bibfnamefont {F.}~\bibnamefont {Kiałka}}, \bibinfo {author}
  {\bibfnamefont {S.}~\bibnamefont {Pedalino}}, \bibinfo {author}
  {\bibfnamefont {M.}~\bibnamefont {Mayor}}, \bibinfo {author} {\bibfnamefont
  {S.}~\bibnamefont {Gerlich}},\ and\ \bibinfo {author} {\bibfnamefont
  {M.}~\bibnamefont {Arndt}},\ }\bibfield  {title} {\bibinfo {title} {Quantum
  superposition of molecules beyond 25 kda},\ }\href
  {https://doi.org/10.1038/s41567-019-0663-9} {\bibfield  {journal} {\bibinfo
  {journal} {Nature Physics}\ }\textbf {\bibinfo {volume} {15}},\ \bibinfo
  {pages} {1242} (\bibinfo {year} {2019})}\BibitemShut {NoStop}%
\bibitem [{\citenamefont {Pedalino}\ \emph {et~al.}(2025)\citenamefont
  {Pedalino}, \citenamefont {Ramírez-Galindo}, \citenamefont {Ferstl},
  \citenamefont {Hornberger}, \citenamefont {Arndt},\ and\ \citenamefont
  {Gerlich}}]{ArndtArxiv}%
  \BibitemOpen
  \bibfield  {author} {\bibinfo {author} {\bibfnamefont {S.}~\bibnamefont
  {Pedalino}}, \bibinfo {author} {\bibfnamefont {B.~E.}\ \bibnamefont
  {Ramírez-Galindo}}, \bibinfo {author} {\bibfnamefont {R.}~\bibnamefont
  {Ferstl}}, \bibinfo {author} {\bibfnamefont {K.}~\bibnamefont {Hornberger}},
  \bibinfo {author} {\bibfnamefont {M.}~\bibnamefont {Arndt}},\ and\ \bibinfo
  {author} {\bibfnamefont {S.}~\bibnamefont {Gerlich}},\ }\href
  {https://doi.org/10.48550/ARXIV.2507.21211} {\bibinfo {title} {Probing
  quantum mechanics using nanoparticle schrödinger cats}},\ \bibinfo
  {howpublished} {arXiv} (\bibinfo {year} {2025}),\ \Eprint
  {https://arxiv.org/abs/2507.21211} {arXiv:2507.21211 [quant-ph]} \BibitemShut
  {NoStop}%
\bibitem [{\citenamefont {Nimmrichter}\ \emph {et~al.}(2025)\citenamefont
  {Nimmrichter}, \citenamefont {R\"atzel}, \citenamefont {Bicket},
  \citenamefont {Seifner},\ and\ \citenamefont {Haslinger}}]{NimmrichterArxiv}%
  \BibitemOpen
  \bibfield  {author} {\bibinfo {author} {\bibfnamefont {S.}~\bibnamefont
  {Nimmrichter}}, \bibinfo {author} {\bibfnamefont {D.}~\bibnamefont
  {R\"atzel}}, \bibinfo {author} {\bibfnamefont {I.~C.}\ \bibnamefont
  {Bicket}}, \bibinfo {author} {\bibfnamefont {M.~S.}\ \bibnamefont
  {Seifner}},\ and\ \bibinfo {author} {\bibfnamefont {P.}~\bibnamefont
  {Haslinger}},\ }\bibfield  {title} {\bibinfo {title} {Electron-enabled
  nanoparticle diffraction},\ }\href {https://doi.org/10.1103/3bvs-ymd7}
  {\bibfield  {journal} {\bibinfo  {journal} {Phys. Rev. Lett.}\ }\textbf
  {\bibinfo {volume} {135}},\ \bibinfo {pages} {173601} (\bibinfo {year}
  {2025})}\BibitemShut {NoStop}%
\bibitem [{\citenamefont {Schattschneider}\ and\ \citenamefont
  {Löffler}(2018)}]{U_v190_i_p39}%
  \BibitemOpen
  \bibfield  {author} {\bibinfo {author} {\bibfnamefont {P.}~\bibnamefont
  {Schattschneider}}\ and\ \bibinfo {author} {\bibfnamefont {S.}~\bibnamefont
  {Löffler}},\ }\bibfield  {title} {\bibinfo {title} {Entanglement and
  decoherence in electron microscopy},\ }\href
  {https://doi.org/10.1016/j.ultramic.2018.04.007} {\bibfield  {journal}
  {\bibinfo  {journal} {Ultramicroscopy}\ }\textbf {\bibinfo {volume} {190}},\
  \bibinfo {pages} {39} (\bibinfo {year} {2018})}\BibitemShut {NoStop}%
\bibitem [{\citenamefont {Schattschneider}\ \emph {et~al.}(2020)\citenamefont
  {Schattschneider}, \citenamefont {Löffler}, \citenamefont {Gollisch},\ and\
  \citenamefont {Feder}}]{JoESaRP_v241_p146810}%
  \BibitemOpen
  \bibfield  {author} {\bibinfo {author} {\bibfnamefont {P.}~\bibnamefont
  {Schattschneider}}, \bibinfo {author} {\bibfnamefont {S.}~\bibnamefont
  {Löffler}}, \bibinfo {author} {\bibfnamefont {H.}~\bibnamefont {Gollisch}},\
  and\ \bibinfo {author} {\bibfnamefont {R.}~\bibnamefont {Feder}},\ }\bibfield
   {title} {\bibinfo {title} {Entanglement and entropy in electron--electron
  scattering},\ }\href {https://doi.org/10.1016/j.elspec.2018.11.009}
  {\bibfield  {journal} {\bibinfo  {journal} {Journal of Electron Spectroscopy
  and Related Phenomena}\ }\textbf {\bibinfo {volume} {241}},\ \bibinfo {pages}
  {146810} (\bibinfo {year} {2020})}\BibitemShut {NoStop}%
\bibitem [{\citenamefont {Blum}(1996)}]{Blum1996}%
  \BibitemOpen
  \bibfield  {author} {\bibinfo {author} {\bibfnamefont {K.}~\bibnamefont
  {Blum}},\ }\href {https://books.google.at/books?id=kl-pMd9Qx04C} {\emph
  {\bibinfo {title} {Density Matrix Theory and Applications}}},\ \bibinfo
  {edition} {2nd}\ ed.,\ Physics of Atoms and Molecules\ (\bibinfo  {publisher}
  {Springer},\ \bibinfo {year} {1996})\BibitemShut {NoStop}%
\bibitem [{\citenamefont {Schattschneider}\ \emph {et~al.}(1999)\citenamefont
  {Schattschneider}, \citenamefont {Nelhiebel},\ and\ \citenamefont
  {Jouffrey}}]{PRB_v59_i16_p10959}%
  \BibitemOpen
  \bibfield  {author} {\bibinfo {author} {\bibfnamefont {P.}~\bibnamefont
  {Schattschneider}}, \bibinfo {author} {\bibfnamefont {M.}~\bibnamefont
  {Nelhiebel}},\ and\ \bibinfo {author} {\bibfnamefont {B.}~\bibnamefont
  {Jouffrey}},\ }\bibfield  {title} {\bibinfo {title} {Density matrix of
  inelastically scattered fast electrons},\ }\href
  {https://doi.org/10.1103/PhysRevB.59.10959} {\bibfield  {journal} {\bibinfo
  {journal} {Phys. Rev. B}\ }\textbf {\bibinfo {volume} {59}},\ \bibinfo
  {pages} {10959} (\bibinfo {year} {1999})}\BibitemShut {NoStop}%
\bibitem [{\citenamefont {Schattschneider}\ \emph {et~al.}(2000)\citenamefont
  {Schattschneider}, \citenamefont {Nelhiebel}, \citenamefont {Souchay},\ and\
  \citenamefont {Jouffrey}}]{M_v31_i4_p333}%
  \BibitemOpen
  \bibfield  {author} {\bibinfo {author} {\bibfnamefont {P.}~\bibnamefont
  {Schattschneider}}, \bibinfo {author} {\bibfnamefont {M.}~\bibnamefont
  {Nelhiebel}}, \bibinfo {author} {\bibfnamefont {H.}~\bibnamefont {Souchay}},\
  and\ \bibinfo {author} {\bibfnamefont {B.}~\bibnamefont {Jouffrey}},\
  }\bibfield  {title} {\bibinfo {title} {The physical significance of the mixed
  dynamic form factor},\ }\href {https://doi.org/10.1016/S0968-4328(99)00112-2}
  {\bibfield  {journal} {\bibinfo  {journal} {Micron}\ }\textbf {\bibinfo
  {volume} {31}},\ \bibinfo {pages} {333 } (\bibinfo {year}
  {2000})}\BibitemShut {NoStop}%
\bibitem [{\citenamefont {Ruimy}\ \emph {et~al.}(2024)\citenamefont {Ruimy},
  \citenamefont {Tziperman}, \citenamefont {Gorlach}, \citenamefont {Mølmer},\
  and\ \citenamefont {Kaminer}}]{nQI_v10_p121}%
  \BibitemOpen
  \bibfield  {author} {\bibinfo {author} {\bibfnamefont {R.}~\bibnamefont
  {Ruimy}}, \bibinfo {author} {\bibfnamefont {O.}~\bibnamefont {Tziperman}},
  \bibinfo {author} {\bibfnamefont {A.}~\bibnamefont {Gorlach}}, \bibinfo
  {author} {\bibfnamefont {K.}~\bibnamefont {Mølmer}},\ and\ \bibinfo {author}
  {\bibfnamefont {I.}~\bibnamefont {Kaminer}},\ }\bibfield  {title} {\bibinfo
  {title} {Many-body entanglement via ‘which-path’ information},\ }\href
  {https://doi.org/10.1038/s41534-024-00899-6} {\bibfield  {journal} {\bibinfo
  {journal} {npj Quantum Information}\ }\textbf {\bibinfo {volume} {10}},\
  \bibinfo {pages} {121} (\bibinfo {year} {2024})}\BibitemShut {NoStop}%
\bibitem [{\citenamefont {Schlosshauer}(2019)}]{PR_v831_p1}%
  \BibitemOpen
  \bibfield  {author} {\bibinfo {author} {\bibfnamefont {M.}~\bibnamefont
  {Schlosshauer}},\ }\bibfield  {title} {\bibinfo {title} {Quantum
  decoherence},\ }\href {https://doi.org/10.1016/j.physrep.2019.10.001}
  {\bibfield  {journal} {\bibinfo  {journal} {Physics Reports}\ }\textbf
  {\bibinfo {volume} {831}},\ \bibinfo {pages} {1} (\bibinfo {year}
  {2019})}\BibitemShut {NoStop}%
\bibitem [{\citenamefont {Delić}\ \emph {et~al.}(2020)\citenamefont {Delić},
  \citenamefont {Reisenbauer}, \citenamefont {Dare}, \citenamefont {Grass},
  \citenamefont {Vuletić}, \citenamefont {Kiesel},\ and\ \citenamefont
  {Aspelmeyer}}]{S_v367_p892}%
  \BibitemOpen
  \bibfield  {author} {\bibinfo {author} {\bibfnamefont {U.}~\bibnamefont
  {Delić}}, \bibinfo {author} {\bibfnamefont {M.}~\bibnamefont {Reisenbauer}},
  \bibinfo {author} {\bibfnamefont {K.}~\bibnamefont {Dare}}, \bibinfo {author}
  {\bibfnamefont {D.}~\bibnamefont {Grass}}, \bibinfo {author} {\bibfnamefont
  {V.}~\bibnamefont {Vuletić}}, \bibinfo {author} {\bibfnamefont
  {N.}~\bibnamefont {Kiesel}},\ and\ \bibinfo {author} {\bibfnamefont
  {M.}~\bibnamefont {Aspelmeyer}},\ }\bibfield  {title} {\bibinfo {title}
  {Cooling of a levitated nanoparticle to the motional quantum ground state},\
  }\href {https://doi.org/10.1126/science.aba3993} {\bibfield  {journal}
  {\bibinfo  {journal} {Science}\ }\textbf {\bibinfo {volume} {367}},\ \bibinfo
  {pages} {892} (\bibinfo {year} {2020})}\BibitemShut {NoStop}%
\bibitem [{\citenamefont {Piotrowski}\ \emph {et~al.}(2023)\citenamefont
  {Piotrowski}, \citenamefont {Windey}, \citenamefont {Vijayan}, \citenamefont
  {Gonzalez-Ballestero}, \citenamefont {de~los Ríos~Sommer}, \citenamefont
  {Meyer}, \citenamefont {Quidant}, \citenamefont {Romero-Isart}, \citenamefont
  {Reimann},\ and\ \citenamefont {Novotny}}]{NP_v19_p1009}%
  \BibitemOpen
  \bibfield  {author} {\bibinfo {author} {\bibfnamefont {J.}~\bibnamefont
  {Piotrowski}}, \bibinfo {author} {\bibfnamefont {D.}~\bibnamefont {Windey}},
  \bibinfo {author} {\bibfnamefont {J.}~\bibnamefont {Vijayan}}, \bibinfo
  {author} {\bibfnamefont {C.}~\bibnamefont {Gonzalez-Ballestero}}, \bibinfo
  {author} {\bibfnamefont {A.}~\bibnamefont {de~los Ríos~Sommer}}, \bibinfo
  {author} {\bibfnamefont {N.}~\bibnamefont {Meyer}}, \bibinfo {author}
  {\bibfnamefont {R.}~\bibnamefont {Quidant}}, \bibinfo {author} {\bibfnamefont
  {O.}~\bibnamefont {Romero-Isart}}, \bibinfo {author} {\bibfnamefont
  {R.}~\bibnamefont {Reimann}},\ and\ \bibinfo {author} {\bibfnamefont
  {L.}~\bibnamefont {Novotny}},\ }\bibfield  {title} {\bibinfo {title}
  {Simultaneous ground-state cooling of two mechanical modes of a levitated
  nanoparticle},\ }\href {https://doi.org/10.1038/s41567-023-01956-1}
  {\bibfield  {journal} {\bibinfo  {journal} {Nature Physics}\ }\textbf
  {\bibinfo {volume} {19}},\ \bibinfo {pages} {1009} (\bibinfo {year}
  {2023})}\BibitemShut {NoStop}%
\bibitem [{\citenamefont {Pontin}\ \emph {et~al.}(2023)\citenamefont {Pontin},
  \citenamefont {Fu}, \citenamefont {Toroš}, \citenamefont {Monteiro},\ and\
  \citenamefont {Barker}}]{NP_v19_p1003}%
  \BibitemOpen
  \bibfield  {author} {\bibinfo {author} {\bibfnamefont {A.}~\bibnamefont
  {Pontin}}, \bibinfo {author} {\bibfnamefont {H.}~\bibnamefont {Fu}}, \bibinfo
  {author} {\bibfnamefont {M.}~\bibnamefont {Toroš}}, \bibinfo {author}
  {\bibfnamefont {T.~S.}\ \bibnamefont {Monteiro}},\ and\ \bibinfo {author}
  {\bibfnamefont {P.~F.}\ \bibnamefont {Barker}},\ }\bibfield  {title}
  {\bibinfo {title} {Simultaneous cavity cooling of all six degrees of freedom
  of a levitated nanoparticle},\ }\href
  {https://doi.org/10.1038/s41567-023-02006-6} {\bibfield  {journal} {\bibinfo
  {journal} {Nature Physics}\ }\textbf {\bibinfo {volume} {19}},\ \bibinfo
  {pages} {1003} (\bibinfo {year} {2023})}\BibitemShut {NoStop}%
\bibitem [{\citenamefont {Kieu}(2023)}]{E_v25_p600}%
  \BibitemOpen
  \bibfield  {author} {\bibinfo {author} {\bibfnamefont {T.~D.}\ \bibnamefont
  {Kieu}},\ }\bibfield  {title} {\bibinfo {title} {Quantum central limit
  theorems, emergence of classicality and time-dependent differential
  entropy},\ }\href {https://doi.org/10.3390/e25040600} {\bibfield  {journal}
  {\bibinfo  {journal} {Entropy}\ }\textbf {\bibinfo {volume} {25}},\ \bibinfo
  {pages} {600} (\bibinfo {year} {2023})}\BibitemShut {NoStop}%
\bibitem [{\citenamefont {Cohen-Tannoudji}(2020)}]{CohenTannoudji2020}%
  \BibitemOpen
  \bibfield  {author} {\bibinfo {author} {\bibfnamefont {C.}~\bibnamefont
  {Cohen-Tannoudji}},\ }\href@noop {} {\emph {\bibinfo {title} {Quantum
  Mechanics}}},\ \bibinfo {edition} {2nd}\ ed.,\ edited by\ \bibinfo {editor}
  {\bibfnamefont {B.}~\bibnamefont {Diu}}, \bibinfo {editor} {\bibfnamefont
  {F.}~\bibnamefont {Laloe}}, \bibinfo {editor} {\bibfnamefont
  {N.}~\bibnamefont {Ostrowsky}},\ and\ \bibinfo {editor} {\bibfnamefont
  {D.}~\bibnamefont {Ostrowsky}}\ (\bibinfo  {publisher} {Wiley-VCH Verlag GmbH
  \& Co. KGaA},\ \bibinfo {address} {Weinheim},\ \bibinfo {year}
  {2020})\BibitemShut {NoStop}%
\bibitem [{\citenamefont {Krivanek}\ \emph {et~al.}(2014)\citenamefont
  {Krivanek}, \citenamefont {Lovejoy}, \citenamefont {Dellby}, \citenamefont
  {Aoki}, \citenamefont {Carpenter}, \citenamefont {Rez}, \citenamefont
  {Soignard}, \citenamefont {Zhu}, \citenamefont {Batson}, \citenamefont
  {Lagos}, \citenamefont {Egerton},\ and\ \citenamefont
  {Crozier}}]{N_v514_p209}%
  \BibitemOpen
  \bibfield  {author} {\bibinfo {author} {\bibfnamefont {O.~L.}\ \bibnamefont
  {Krivanek}}, \bibinfo {author} {\bibfnamefont {T.~C.}\ \bibnamefont
  {Lovejoy}}, \bibinfo {author} {\bibfnamefont {N.}~\bibnamefont {Dellby}},
  \bibinfo {author} {\bibfnamefont {T.}~\bibnamefont {Aoki}}, \bibinfo {author}
  {\bibfnamefont {R.~W.}\ \bibnamefont {Carpenter}}, \bibinfo {author}
  {\bibfnamefont {P.}~\bibnamefont {Rez}}, \bibinfo {author} {\bibfnamefont
  {E.}~\bibnamefont {Soignard}}, \bibinfo {author} {\bibfnamefont
  {J.}~\bibnamefont {Zhu}}, \bibinfo {author} {\bibfnamefont {P.~E.}\
  \bibnamefont {Batson}}, \bibinfo {author} {\bibfnamefont {M.~J.}\
  \bibnamefont {Lagos}}, \bibinfo {author} {\bibfnamefont {R.~F.}\ \bibnamefont
  {Egerton}},\ and\ \bibinfo {author} {\bibfnamefont {P.~A.}\ \bibnamefont
  {Crozier}},\ }\bibfield  {title} {\bibinfo {title} {Vibrational spectroscopy
  in the electron microscope},\ }\href {https://doi.org/10.1038/nature13870}
  {\bibfield  {journal} {\bibinfo  {journal} {Nature}\ }\textbf {\bibinfo
  {volume} {514}},\ \bibinfo {pages} {209} (\bibinfo {year}
  {2014})}\BibitemShut {NoStop}%
\bibitem [{\citenamefont {Kirkland}(1998)}]{Kirkland1998}%
  \BibitemOpen
  \bibfield  {author} {\bibinfo {author} {\bibfnamefont {E.~J.}\ \bibnamefont
  {Kirkland}},\ }\href {https://doi.org/10.1007/978-1-4757-4406-4} {\emph
  {\bibinfo {title} {Advanced computing in electron microscopy}}}\ (\bibinfo
  {publisher} {Plenum Press},\ \bibinfo {year} {1998})\BibitemShut {NoStop}%
\bibitem [{\citenamefont {Löffler}\ \emph {et~al.}(2013)\citenamefont
  {Löffler}, \citenamefont {Motsch},\ and\ \citenamefont
  {Schattschneider}}]{U_v131_i0_p39}%
  \BibitemOpen
  \bibfield  {author} {\bibinfo {author} {\bibfnamefont {S.}~\bibnamefont
  {Löffler}}, \bibinfo {author} {\bibfnamefont {V.}~\bibnamefont {Motsch}},\
  and\ \bibinfo {author} {\bibfnamefont {P.}~\bibnamefont {Schattschneider}},\
  }\bibfield  {title} {\bibinfo {title} {A pure state decomposition approach of
  the mixed dynamic form factor for mapping atomic orbitals},\ }\href
  {https://doi.org/10.1016/j.ultramic.2013.03.021} {\bibfield  {journal}
  {\bibinfo  {journal} {Ultramicroscopy}\ }\textbf {\bibinfo {volume} {131}},\
  \bibinfo {pages} {39} (\bibinfo {year} {2013})},\ \Eprint
  {https://arxiv.org/abs/1210.2947} {1210.2947} \BibitemShut {NoStop}%
\bibitem [{\citenamefont {Rusz}\ \emph {et~al.}(2017)\citenamefont {Rusz},
  \citenamefont {Lubk}, \citenamefont {Spiegelberg},\ and\ \citenamefont
  {Tyutyunnikov}}]{PRB_v96_p245121a}%
  \BibitemOpen
  \bibfield  {author} {\bibinfo {author} {\bibfnamefont {J.}~\bibnamefont
  {Rusz}}, \bibinfo {author} {\bibfnamefont {A.}~\bibnamefont {Lubk}}, \bibinfo
  {author} {\bibfnamefont {J.}~\bibnamefont {Spiegelberg}},\ and\ \bibinfo
  {author} {\bibfnamefont {D.}~\bibnamefont {Tyutyunnikov}},\ }\bibfield
  {title} {\bibinfo {title} {Fully nonlocal inelastic scattering computations
  for spectroscopical transmission electron microscopy methods},\ }\href
  {https://doi.org/10.1103/physrevb.96.245121} {\bibfield  {journal} {\bibinfo
  {journal} {Physical Review B}\ }\textbf {\bibinfo {volume} {96}},\ \bibinfo
  {pages} {245121} (\bibinfo {year} {2017})}\BibitemShut {NoStop}%
\bibitem [{\citenamefont {Bateman}\ \emph {et~al.}(2014)\citenamefont
  {Bateman}, \citenamefont {Nimmrichter}, \citenamefont {Hornberger},\ and\
  \citenamefont {Ulbricht}}]{Bateman2014}%
  \BibitemOpen
  \bibfield  {author} {\bibinfo {author} {\bibfnamefont {J.~E.}\ \bibnamefont
  {Bateman}}, \bibinfo {author} {\bibfnamefont {S.}~\bibnamefont
  {Nimmrichter}}, \bibinfo {author} {\bibfnamefont {K.}~\bibnamefont
  {Hornberger}},\ and\ \bibinfo {author} {\bibfnamefont {H.}~\bibnamefont
  {Ulbricht}},\ }\bibfield  {title} {\bibinfo {title} {Near-field
  interferometry of a free-falling nanoparticle from a point-like source},\
  }\href {https://doi.org/10.1038/ncomms5788} {\bibfield  {journal} {\bibinfo
  {journal} {Nature comm.}\ }\textbf {\bibinfo {volume} {5}} (\bibinfo {year}
  {2014})}\BibitemShut {NoStop}%
\bibitem [{\citenamefont {Romero-Isart}\ \emph {et~al.}(2011)\citenamefont
  {Romero-Isart}, \citenamefont {Pflanzer}, \citenamefont {Blaser},
  \citenamefont {Kaltenbaek}, \citenamefont {Kiesel}, \citenamefont
  {Aspelmeyer},\ and\ \citenamefont {Cirac}}]{PRL_v107_p20405}%
  \BibitemOpen
  \bibfield  {author} {\bibinfo {author} {\bibfnamefont {O.}~\bibnamefont
  {Romero-Isart}}, \bibinfo {author} {\bibfnamefont {A.~C.}\ \bibnamefont
  {Pflanzer}}, \bibinfo {author} {\bibfnamefont {F.}~\bibnamefont {Blaser}},
  \bibinfo {author} {\bibfnamefont {R.}~\bibnamefont {Kaltenbaek}}, \bibinfo
  {author} {\bibfnamefont {N.}~\bibnamefont {Kiesel}}, \bibinfo {author}
  {\bibfnamefont {M.}~\bibnamefont {Aspelmeyer}},\ and\ \bibinfo {author}
  {\bibfnamefont {J.~I.}\ \bibnamefont {Cirac}},\ }\bibfield  {title} {\bibinfo
  {title} {Large quantum superpositions and interference of massive
  nanometer-sized objects},\ }\href
  {https://doi.org/10.1103/physrevlett.107.020405} {\bibfield  {journal}
  {\bibinfo  {journal} {Physical Review Letters}\ }\textbf {\bibinfo {volume}
  {107}},\ \bibinfo {pages} {020405} (\bibinfo {year} {2011})}\BibitemShut
  {NoStop}%
\bibitem [{\citenamefont {Rossi}\ \emph {et~al.}(2024)\citenamefont {Rossi},
  \citenamefont {Militaru}, \citenamefont {Zambon}, \citenamefont
  {Riera-Campeny}, \citenamefont {Romero-Isart}, \citenamefont {Frimmer},\ and\
  \citenamefont {Novotny}}]{Rossi2024}%
  \BibitemOpen
  \bibfield  {author} {\bibinfo {author} {\bibfnamefont {M.}~\bibnamefont
  {Rossi}}, \bibinfo {author} {\bibfnamefont {A.}~\bibnamefont {Militaru}},
  \bibinfo {author} {\bibfnamefont {N.~C.}\ \bibnamefont {Zambon}}, \bibinfo
  {author} {\bibfnamefont {A.}~\bibnamefont {Riera-Campeny}}, \bibinfo {author}
  {\bibfnamefont {O.}~\bibnamefont {Romero-Isart}}, \bibinfo {author}
  {\bibfnamefont {M.}~\bibnamefont {Frimmer}},\ and\ \bibinfo {author}
  {\bibfnamefont {L.}~\bibnamefont {Novotny}},\ }\href
  {https://doi.org/10.48550/ARXIV.2408.01264} {\bibinfo {title} {Quantum
  delocalization of a levitated nanoparticle}} (\bibinfo {year}
  {2024})\BibitemShut {NoStop}%
\bibitem [{\citenamefont {Tebbenjohanns}\ \emph {et~al.}(2021)\citenamefont
  {Tebbenjohanns}, \citenamefont {Mattana}, \citenamefont {Rossi},
  \citenamefont {Frimmer},\ and\ \citenamefont {Novotny}}]{N_v595_p378}%
  \BibitemOpen
  \bibfield  {author} {\bibinfo {author} {\bibfnamefont {F.}~\bibnamefont
  {Tebbenjohanns}}, \bibinfo {author} {\bibfnamefont {M.~L.}\ \bibnamefont
  {Mattana}}, \bibinfo {author} {\bibfnamefont {M.}~\bibnamefont {Rossi}},
  \bibinfo {author} {\bibfnamefont {M.}~\bibnamefont {Frimmer}},\ and\ \bibinfo
  {author} {\bibfnamefont {L.}~\bibnamefont {Novotny}},\ }\bibfield  {title}
  {\bibinfo {title} {Quantum control of a nanoparticle optically levitated in
  cryogenic free space},\ }\href {https://doi.org/10.1038/s41586-021-03617-w}
  {\bibfield  {journal} {\bibinfo  {journal} {Nature}\ }\textbf {\bibinfo
  {volume} {595}},\ \bibinfo {pages} {378} (\bibinfo {year}
  {2021})}\BibitemShut {NoStop}%
\bibitem [{\citenamefont {Dania}\ \emph {et~al.}(2025)\citenamefont {Dania},
  \citenamefont {Kremer}, \citenamefont {Piotrowski}, \citenamefont {Candoli},
  \citenamefont {Vijayan}, \citenamefont {Romero-Isart}, \citenamefont
  {Gonzalez-Ballestero}, \citenamefont {Novotny},\ and\ \citenamefont
  {Frimmer}}]{NP_v21_p1603}%
  \BibitemOpen
  \bibfield  {author} {\bibinfo {author} {\bibfnamefont {L.}~\bibnamefont
  {Dania}}, \bibinfo {author} {\bibfnamefont {O.~S.}\ \bibnamefont {Kremer}},
  \bibinfo {author} {\bibfnamefont {J.}~\bibnamefont {Piotrowski}}, \bibinfo
  {author} {\bibfnamefont {D.}~\bibnamefont {Candoli}}, \bibinfo {author}
  {\bibfnamefont {J.}~\bibnamefont {Vijayan}}, \bibinfo {author} {\bibfnamefont
  {O.}~\bibnamefont {Romero-Isart}}, \bibinfo {author} {\bibfnamefont
  {C.}~\bibnamefont {Gonzalez-Ballestero}}, \bibinfo {author} {\bibfnamefont
  {L.}~\bibnamefont {Novotny}},\ and\ \bibinfo {author} {\bibfnamefont
  {M.}~\bibnamefont {Frimmer}},\ }\bibfield  {title} {\bibinfo {title}
  {High-purity quantum optomechanics at room temperature},\ }\href
  {https://doi.org/10.1038/s41567-025-02976-9} {\bibfield  {journal} {\bibinfo
  {journal} {Nature Physics}\ }\textbf {\bibinfo {volume} {21}},\ \bibinfo
  {pages} {1603} (\bibinfo {year} {2025})}\BibitemShut {NoStop}%
\bibitem [{\citenamefont {Troyer}\ \emph {et~al.}(2025)\citenamefont {Troyer},
  \citenamefont {Fechtel}, \citenamefont {Hummer}, \citenamefont {Rudolph},
  \citenamefont {Stickler}, \citenamefont {Delić},\ and\ \citenamefont
  {Arndt}}]{ArndtArxiv2}%
  \BibitemOpen
  \bibfield  {author} {\bibinfo {author} {\bibfnamefont {S.}~\bibnamefont
  {Troyer}}, \bibinfo {author} {\bibfnamefont {F.}~\bibnamefont {Fechtel}},
  \bibinfo {author} {\bibfnamefont {L.}~\bibnamefont {Hummer}}, \bibinfo
  {author} {\bibfnamefont {H.}~\bibnamefont {Rudolph}}, \bibinfo {author}
  {\bibfnamefont {B.~A.}\ \bibnamefont {Stickler}}, \bibinfo {author}
  {\bibfnamefont {U.}~\bibnamefont {Delić}},\ and\ \bibinfo {author}
  {\bibfnamefont {M.}~\bibnamefont {Arndt}},\ }\href
  {https://doi.org/10.48550/ARXIV.2509.13398} {\bibinfo {title} {Quantum
  ground-state cooling of two librational modes of a nanorotor}},\ \bibinfo
  {howpublished} {arXiv} (\bibinfo {year} {2025}),\ \Eprint
  {https://arxiv.org/abs/2509.13398} {arXiv:2509.13398 [quant-ph]} \BibitemShut
  {NoStop}%
\bibitem [{Note1()}]{Note1}%
  \BibitemOpen
  \bibinfo {note} {\protect \url
  {https://en.wikipedia.org/wiki/Characteristic_function_(probability_theory)},
  accessed March 6th, 2026}\BibitemShut {NoStop}%
\end{thebibliography}
\end{document}